\documentstyle[10pt,fleqn]{article}
\begin{document}

\hfill{\sl preprint - Imperial/TP/96-97/37, hep-ph/9703323}
\par\bigskip\par\rm
\par\bigskip\begin{center}\LARGE

{\bf Inclusion of Chemical Potential for Scalar Fields}

\end{center}\par\bigskip\par\rm\normalsize

\begin{center}\Large Antonio Filippi 
\footnote{e-mail: \sl a.filippi@ic.ac.uk\rm}\end{center}

\begin{center}\large\it{ 
Theoretical Physics Group, Imperial College, \\
Prince Consort Road, London SW7 2BZ, U.K.} 
\end{center}\rm\normalsize

\par\bigskip\par\sl\hfill 15 March 1997 \par\medskip\par\rm

\hrule\par\begin{description}\item{Abstract: }\it \\
There are two possible methods for the inclusion of the chemical potential in a
relativistic bosonic field theory. The most popular method has recently been
criticised by some authors, so much so as to require a re-analysis of the
entire problem. I here present a new approach to the inclusion of the chemical
potential that uses the second method and zeta-function regularisation
techniques. I first apply it to the non-interacting field, obtaining an
expression for the effective potential which is formally coincident with the
well-known one from the first method. My approach, however, seems here
mathematically more rigorous since it excludes the presence of the
multiplicative anomaly. I then obtain a new expression for the one loop
effective potential for the interacting field. This is regularised, presents
less cryptic excitation energies, shows Goldstone excitations, and in the
non-interacting limit gives the expected expression. The high-temperature
expansion is calculated easily and with clarity, and confirms the results of
the first method for the critical temperature.
                                                                            
\par\end{description}\hrule\par\medskip\rm

\smallskip\noindent{\sl PACS numbers:
\hspace{0.2cm} 11.10.Wx, 05.30.Jp, 11.30.Qc}\par\bigskip\rm


\section{Introduction}\renewcommand{\theequation}
{\mbox{\arabic{section}.\arabic{equation}}}\setcounter{equation}{0}

The calculation of the one loop effective potential for a relativistic
self-interacting bosonic field in the presence of a net global charge has been
the subject of interest for a long time.
The first calculations for the inclusion of the chemical potential in a
relativistic field theory were done by Kapusta \cite{kap1}\cite{kap} and by
Haber and Weldon \cite{hw1}, in 1981. Kapusta calculated the partition function
for a {\em non-interacting} charged scalar field, showing the presence of
relativistic Bose-Einstein condensation. He also showed, using an
approximation, that in the self-interacting case the chemical potential can
modify the phase transition. Haber and Weldon confirmed these results and
calculated, amongst other things, the high-temperature expansion of the
thermodynamic potential \cite{hw2}\cite{hw3}.
 
These calculations, in the free field case, were later also developed using
zeta-function and heat-kernel regularisation techniques
\cite{haw}\cite{act}\cite{rew} and extended to curved spaces
\cite{dowsch}\cite{kir}\cite{cogvan}.
 
It was only recently though, with the work of Benson, Bernstein and Dodelson
\cite{bd}\cite{bd2}, that an exact expression was obtained for the one loop
effective potential for a {\em self-interacting} charged scalar field. These
authors included the chemical potential with the method, already used by
Kapusta and by Haber and Weldon, and which I will call Method I, of inserting
in the functional integrals that determine the thermodynamic partition
function, an effective Hamiltonian $H-\mu Q$, where $H$ is the microscopic
Hamiltonian, $Q$ the charge operator and $\mu$ the chemical potential. Having
calculated the effective potential, they demonstrated, in the high-temperature
limit, how the chemical potential noticeably changes the phase structure of the
system, and confirmed the interpretation of spontaneous symmetry breaking as
Bose-Einstein condensation.

This remarkable work, however, has some unclear aspects. In particular, it is
the approach itself for including the chemical potential that is criticised by
some authors.
Evans' analysis \cite{evans} shows how Method I is ``unphysical'' in the sense
that it is not clear in the distinction between micro and macroscopic
conditions, between evolution and boundary conditions and in the definition of
the energies.
There is also the problem of the definition of the charge operator in the case
of broken symmetry. According to the considerations of Chaichian, Montonen,
P\'erez Rojas and others, presented in a recent series of articles
\cite{chai1}\cite{chai2}\cite{chai3}, it is not legitimate to use, in the
time-evolution operator, the charge of a symmetry that can be broken, because
in these conditions the charge is not defined \cite{orzalesi}\cite{strocchi}. 
Benson, Bernstein and Dodelson's effective potential is then obtained by formal
manipulation of divergent terms and is also not clear in defining the energies
of the particles. For this latter reason Benson, Bernstein and Dodelson can
only study the symmetry breaking in the high-temperature limit which is
obtained in a decidedly complicated way.
Method I, moreover, as I will show later, in the case of interacting field does
not easily lend itself to regularisation with the zeta function and, in
literature, it is necessary to resort to a heat-kernel expansion \cite{kirtom}.
There remain also some doubts raised by Chaichian and colleagues \cite{chai3}
concerning the renormalization in the presence of chemical potential carried
out by Benson, Bernstein and Dodelson. These could be resolved by studying the
Seeley-De Witt coefficients once an approach with the zeta function is
possible.
 
Besides the method of including the chemical potential used by Benson,
Bernstein and Dodelson and those who preceded them, there exists another one
which, as Evans recently observed \cite{evans}, distinguishes in a much clearer
way between microscopic and macroscopic quantities, keeping separate in the
formalism the interaction Hamiltonian $H$ and the chemical potential. Very
little has been calculated with this method \cite{lanwee}\cite{kosewe}, and it
has never been used to calculate the effective potential. Following Evans'
example I will refer to this alternative as Method II.

{}From what was said regarding Method I it is clear that it is necessary to
develop an alternative approach to the calculation of the effective potential,
an approach which does not suffer from the same ambiguities of the existing
one, that avoids, for example, the need to manipulate infinities, that permits
the use of regularisation techniques, bypasses the problem of the charge
operator and provides less cryptic excitation energies. {\em The development of
such an approach goes beyond being interesting, it is fundamental for verifying
whether it is  actually possible to include the chemical potential in an
interacting field theory}.  

The aim of this work is the critical comparison of the two methods, the
analysis of the unclear aspects or of the difficulties in the practical use of
both, but most of all the development of a new approach to the calculation of
the effective potential, based on Method II and the zeta-function and
heat-kernel regularisation techniques. 

In this work I will make extensive use of zeta-function regularisation, a
technique introduced by Dowker and Critchley \cite{dowcri} and Hawking
\cite{haw}, which is particularly useful when dealing with the ultraviolet
divergences arising in the calculation of the one loop effective potential
\cite{mono}. This is related to the small disturbances operator $A$ (i.e. the
operator extracted by the quadratic term in the expanded action) by:
\begin{equation}
V_{eff}=\frac{S^{0}}{\beta V}+\frac{1}{2\beta V}\ln\det(A)
\end{equation}
where $\beta$ is the inverse of the equilibrium temperature of the system.
This determinant, as a product of eigenvalues, is formally divergent. The
zeta-function regularisation gives it a meaning in the sense of analytical
continuation.
The zeta function $\zeta(s|A)$ for a small disturbances operator $A$ can be
explicitly expressed starting from the Mellin transform of the trace of the
heat operator related to $A$:
\begin{equation}
\zeta(s|A)=\frac{1}{\Gamma (s)}\int_{0}^{\infty} t^{s-1}Tr\,e^{-tA}dt
\end{equation}
This function can be analytically continued to $s=0$, so that its derivative is
well defined. It can be shown that the regularised one loop effective potential
reads:
\begin{equation}
V_{eff}=\frac{S^{0}}{\beta V}-\frac{\zeta'(0|A)}{2\beta V}+
\frac{\zeta(0|A)}{2\beta V}\ln(\ell^{2})
\end{equation}
where $\ell^{2}$ is a renormalization scale.

In section 2 I will briefly introduce the two methods of including the chemical
potential. In section 3 I will present a new approach to the calculation of the
effective potential applying it to the case of a non-interacting field and
showing how it produces the same formal expression for the effective potential
as the more classical method, but in a mathematically more rigorous way. In
section 4 I will show the propagators obtained with the two approaches and will
then re-propose a comparison of the two methods done by Evans. In section 5 I
will briefly present the standard approach to the interacting case, analysing
some of its weaknesses. In section 6 I will finally develop my approach to this
case, obtaining a new expression for the one loop effective potential and
analysing various aspects of it. 

\section{The two methods}\renewcommand{\theequation}
{\mbox{\arabic{section}.\arabic{equation}}}\setcounter{equation}{0}

I am interested in the calculation of the one loop effective potential for a
theory described by the microscopic Hamiltonian density:
\begin{equation}
{\cal H} = \frac{1}{2}\pi_{i}\pi_{i}+\frac{1}{2}\vec{\nabla}\phi_{i}
\vec{\nabla}\phi_{i}+\frac{1}{2}m^{2}\phi^{2}+\frac{\lambda}{4!}\phi^{4}
\label{hamint}
\end{equation}
where $\phi^{2} = \phi^{2}_{1}+\phi^{2}_{2}$ and the sum over the repeated
indices  $i=1,2$ is understood.
This Hamiltonian is invariant under a global O(2) symmetry. Noether's theorem
then guarantees the existence of a conserved current $J_{\nu}$  and the
correspondent charge. The latter is {\em formally} defined as the space
integral of the current $J_{0}$ and is a constant of motion, and also a
generator of the symmetry. In our case the charge density can be easily derived
as:
\begin{equation}
{\cal Q} = \phi_{1}\pi_{2}-\phi_{2}\pi_{1}
\end{equation}
We therefore have a quantum operator $Q$ associated with an additive quantity
of the system and can define the grandcanonical partition function:
\begin{equation}
 Z  = Tr(e^{-\beta(H-\mu Q)}) = \int d\phi <\!\!\phi|
e^{-\beta (H-\mu Q)}|\phi\!\!>\label{Z=intdephi2} 
\end{equation}

It is now a matter of finding useful expressions for the calculation of this
partition function, using functional integrals.

In general:
\begin{equation}
Z = N\int_{BC}[d\phi_{1}] [d\phi_{2}] e^{-S} \label{ealla-s}
\end{equation}
where BC are boundary conditions on the fields and S the classical action that
is obtained from the chosen Hamiltonian.

For Method I:
\begin{eqnarray}
S = \int_{0}^{\beta}d\tau\int d^{3}x\left[\frac{1}{2}\left[(
\partial_{\tau}\phi_{1})^{2}\right.\right.\!\!\!\!\! &+&\!\!\!\!
\left. (\partial_{\tau}\phi_{2})^{2}+(\vec{\nabla}\phi_{1})^{2}+
(\vec{\nabla}\phi_{2})^{2}+m^{2}\phi^{2}\right]+
\frac{\lambda}{4!}\phi^{4}+\nonumber\\ 
        &+&\!\!\!\!\left.i\mu(\phi_{2}\partial_{\tau}\phi_{1}-
\phi_{1}\partial_{\tau}\phi_{2})-\frac{1}{2}\mu^{2}\phi^{2}\right] 
\label{azioneI}
\end{eqnarray}
I have emphasised the term in $\mu^{2}$, which appears only due to the presence
of the charge density ${\cal Q}$ in the effective Hamiltonian. This is a very
important term for the development of the discussion. We can get an idea of
this by observing that this action presents an ``effective mass'' term
$m^{2}-\mu^{2}$. We expect therefore that this new contribution itself gives
symmetry breaking when $\mu^{2}$ is greater than $m^{2}$, even for positive
$m^{2}$.

The boundary conditions on the fields are the well-known periodic ones:
\begin{equation} 
  \phi(\tau)=\phi(\tau+\beta)
\end{equation} 

This approach does of course present some advantages. Including the chemical
potential in the Hamiltonian, other than being intuitive, allows us to have an
analogous theory to that without net charge and the periodic boundary
conditions simplify the calculations.
It is therefore understandable that this approach is more widely used, if not
the only one used, for practical calculations
\cite{kap1}\cite{kap}\cite{hw1}\cite{bd}\cite{bd2}.
 
For the less well-known Method II instead, one obtains:
\begin{equation}
S = \int_{0}^{\beta}d\tau\int d^{3}x\left[\frac{1}{2}
\left[(\partial_{\tau}\phi_{1})^{2}+ (\partial_{\tau}\phi_{2})^{2}+
(\vec{\nabla}\phi_{1})^{2}+(\vec{\nabla}\phi_{2})^{2}+m^{2}\phi^{2}\right]+
\frac{\lambda}{4!}\phi^{4}\right] \label{azioneII}
\end{equation}
Now the chemical potential is not present in the action and only the true
microscopic Lagrangian appears.

Instead of the usual periodic conditions, we have {\em pseudoperiodic
conditions} on the fields, that contain the chemical potential
\cite{lanwee}\cite{kosewe}:
\begin{equation}
\phi(\tau)=e^{\beta\mu}\phi(\tau+\beta)
\end{equation}
These boundary conditions are less convenient than the periodic ones and it is
probably this that has prevented a greater use of this method but, as I will
show later, this is balanced by much simpler small disturbances operators. 

\section{The non-interacting field}\renewcommand{\theequation}
{\mbox{\arabic{section}.\arabic{equation}}}\setcounter{equation}{0}

The case of a non-interacting field has been widely studied with Method I. It
remains interesting, though, for understanding where the characteristic
problems of the interacting case arise. It is also an important test bench for
any attempt at a new approach to the problem.

With regards to Method I, it can be easily seen from \ref{azioneI} that with
the non-interacting action is associated an operator $A_{I}$: 
\begin{equation}
\left(\begin{array}{cc} {-\partial_{\tau}^{2}-\vec{\nabla}^{2}+m^{2}-
\mu^{2}} & {-2i\mu\partial_{\tau}} \\ {2i\mu\partial_{\tau}} & 
{-\partial_{\tau}^{2}-\vec{\nabla}^{2}+m^{2}-\mu^{2}} \end{array} \right)
\end{equation}
This is a matrix valued elliptic differential operator, but not self-adjoint,
due to the presence of the chemical potential. 

I am interested in the effective potential, which is linked to this operator
by:
\begin{equation}
V_{eff}=\frac{1}{2\beta V}\ln\det(A_{I})
\end{equation}
As $A_{I}$ is matrix valued we {\em formally} have:
\begin{eqnarray}
V_{eff}&=&\frac{1}{2\beta
V}\ln\det[(-\partial_{\tau}^{2}-\vec{\nabla}^{2}+
m^{2}-\mu^{2})^{2}-4\mu^{2}\partial_{\tau}^{2}]=\nonumber \\
       &=&\frac{1}{2\beta V}\ln\det\{[(i\partial_{\tau}-i\mu)^{2}-
\vec{\nabla}^{2}+m^{2}][(i\partial_{\tau}+i\mu)^{2}-\vec{\nabla}^{2}+
m^{2}]\}=\nonumber \\
       &=&\frac{1}{2\beta V}\ln\det[(i\partial_{\tau}-i\mu)^{2}-
\vec{\nabla}^{2}+m^{2}]+\frac{1}{2\beta
V}\ln\det[(i\partial_{\tau}+i\mu)^{2}-
\vec{\nabla}^{2}+m^{2}] \label{loganom}
\end{eqnarray}
It is now easy to find the eigenvalues of these operators as:
\begin{equation}
(\omega_{n}\pm i\mu)^{2}+k^{2}+m^{2}
\end{equation}
with $\omega_{n}=\frac{2\pi}{\beta}n$.

With these eigenvalues:
\begin{equation}
Tr\,e^{-tA_{I}}=V\sum_{n}\int\frac{d^{3}k}{(2\pi)^{3}}
[e^{-t[(\omega_{n}+
i\mu)^{2}+k^{2}+m^{2}]}+e^{-t[(\omega_{n}-i\mu)^{2}+k^{2}+m^{2}]}]
\end{equation}
{}From this trace of the heat kernel the zeta function can be easily calculated
and hence also the effective potential.

Infact attention has only recently been drawn \cite{anomalia} to the fact that
two zeta-function regularised determinants $\det(B)$ e $\det(C)$ do not
satisfy, in general, a relationship such as $\ln\det(BC)=\ln\det(B)+\ln\det(C)$
(\ref{loganom}) and to assume its validity (as has always been done up to now)
can cause a quantity called multiplicative anomaly to be neglected. This
quantity is given by:
\begin{equation}
\ln F(BC)=\ln\det(BC)-\ln\det(B)-\ln\det(C)
\end{equation}
Such a quantity may be found to be physically relevant, but it seems not easily
computable, especially for not self-adjoint operators like those with which we
are working.

We will now look at how it is possible to calculate the non-interacting
effective potential with a new approach that uses Method II and the
zeta-function regularisation techniques.

To the action of \ref{azioneII}, for $\lambda=0$, is associated an operator
$A_{II}$:
\begin{equation}
\left(\begin{array}{cc} {-\partial_{\tau}^{2}-\vec{\nabla}^{2}+m^{2}}
& 
{0} \\ {0} & {-\partial_{\tau}^{2}-\vec{\nabla}^{2}+m^{2}} \end{array}
\right)
 \label{AIIlibero}
\end{equation}
This is a matrix valued elliptic differential operator, not self-adjoint.
An important characteristic for us, with respect to the corresponding operator
calculated with Method I, is that here there are no differential terms on the
off-diagonal. As we will see later, this will greatly simplify the calculations
with the zeta function in the self-interacting case.
 
My approach also has another remarkable advantage with respect to Method I: the
operator that has just been calculated is made up, on its diagonal, of two
identical operatorial terms and therefore does not present any multiplicative
anomaly.

Even if I have an operator which is not self-adjoint, but is elliptic, I can
still define the $\log\det (A_{II})$ via the zeta function and I can calculate
it explicitly as I know its spectrum. 
Even though the chemical potential does not appear explicitly in the operator,
because of the boundary conditions it appears in the couples of eigenvalues:
\begin{equation}
(\omega_{n}+i\mu)^{2}+k^{2}+m^{2}
\end{equation}
The trace of the heat kernel will therefore be:
\begin{eqnarray}
Tr\,e^{-tA_{II}}&=&V\int\frac{d^{3}k}{(2\pi)^{3}} \sum_{n} 
\left[e^{-t[(\omega_{n}+i\mu)^{2}+k^{2}+m^{2}]}+e^{-t[(\omega_{n}+i\mu)^{2}+
k^{2}+m^{2}]}\right] = \nonumber \\
               &=&2 V\int\frac{d^{3}k}{(2\pi)^{3}} \sum_{n}
e^{-t[(\omega_{n}+
i\mu)^{2}+k^{2}+m^{2}]} \label{tracciadue}
\end{eqnarray}
It is therefore possible to continue with the standard methods of zeta-function
regularisation and obtain the well-known:
\begin{eqnarray}
V_{eff}&=&-\frac{1}{16\pi^{2}}\frac{m^{4}}{2}\left[-\ln(m^{2}\ell^{2})+
\frac{3}{2}\right]+\nonumber \\
       &+& \frac{1}{\beta}\int\frac{d^{3}k}{(2\pi)^{3}}  
\ln(1-e^{-\beta(\sqrt{k^{2}+m^{2}}-\mu)})+\frac{1}{\beta} 
\int\frac{d^{3}k}{(2\pi)^{3}}
\ln(1-e^{-\beta(\sqrt{k^{2}+m^{2}}+\mu)})
\nonumber \\ \label{well-known}
\end{eqnarray}
This expression coincides with the analogous ones calculated with Method I but,
unlike those, it does not require any eventual additional terms due to the
anomaly.

It is now easy to see how the two approaches anyhow coincide except for the
possible multiplicative anomaly.

The eigenvalues of the two operators are different, but what is important for
us is the trace of the heat kernel.

With Method I:
\begin{equation}
Tr\,e^{-tA_{I}}=V\int\frac{d^{3}k}{(2\pi)^{3}} \sum_{n} 
\left[e^{-t[(\omega_{n}+i\mu)^{2}+k^{2}+m^{2}]}+e^{-t[(\omega_{n}-i\mu)^{2}
+k^{2}+m^{2}]}\right]
\end{equation}
but:
\begin{equation}
 (\omega_{-n} - i\mu)^{2}+k^{2}+m^{2} = (-\frac{2\pi}{\beta}n-
i\mu)^{2}+k^{2}+
m^{2}=(\omega_{n} + i\mu)^{2}+k^{2}+m^{2}
\end{equation}
and the summation is for all the integer numbers.

Hence, also for Method I, the trace of the heat kernel (except for the anomaly)
is:
\begin{equation}
Tr\,e^{-tA_{I}}=2V\int\frac{d^{3}k}{(2\pi)^{3}} \sum_{n}
e^{-t[(\omega_{n}+
i\mu)^{2}+k^{2}+m^{2}]}=Tr\,e^{-tA_{II}}
\end{equation}
Once the trace of the heat kernel has been calculated, via a Mellin transform
one obtains the zeta function and from this the effective potential. Because
this is the only object from which we can obtain physical information about the
system, we can see how the two approaches, in the charged free case, give
coinciding results (except for the multiplicative anomaly), my approach being
more rigorous and reliable if the anomaly is found to be influential in this
case. 

\section{Propagators}\renewcommand{\theequation}
{\mbox{\arabic{section}.\arabic{equation}}}\setcounter{equation}{0}

An analysis of what I have introduced so far can show how Method I is lacking
at least in physical clarity. As we have seen, infact, Method II keeps that
which determines the microscopic interactions, the Hamiltonian of the
zero-temperature theory, well separate from that which determines the
``macroscopic'' boundary conditions, the chemical potential. We have, infact,
taken as the complex time-evolution operator the term $e^{-\tau H}$ and only
with this have we established the relationship between partition function and
functional integrals. The chemical potential, instead, only appears in the
boundary conditions. Method I, on the contrary, inserts the macroscopic
information together with the microscopic ones directly into an effective
Hamiltonian which it then uses both for the thermodynamics as well as for the
microscopic evolution of the system, with a complex time-evolution operator
$e^{-\tau (H-\mu Q)}$. 

The problem is not just about ``aesthetic'' differences or, as we also shall
see, of greater or lesser ease of calculations. In the near-total absence of
material in literature, a recent comparison was made by Evans
\cite{evans}. Evans gives substance to these observations with a comparison of
the propagators, calculated with both methods.

It is possible to calculate the thermal propagator (euclidean signature) also
with my approach by integrating on the parameter $t$ in the heat kernel
associated with the operator $A_{II}$. 

I here show only the result, which is analogous with that already present in
literature and calculated with more classical methods
\cite{evans}\cite{lanwee}.
\begin{eqnarray}
G(x,0)&=&\int\frac{dk_{0}}{2\pi}\int\frac{d^{3}k}{(2\pi)^{3}}  
\frac{e^{i
\vec{k}\vec{x}+ik_{0}\tau}}{k_{0}^{2}+\omega_{\!\vec{\kappa}}^{2}}+
\nonumber \\
      &+&\int\frac{d^{3}k}{(2\pi)^{3}}  e^{i
\vec{k}\vec{x}}\left(\frac{1}
{e^{\beta(\omega_{\!\vec{\kappa}}-\mu)}-1}\frac{e^{-\omega_{\!\vec{\kappa}}
\tau}}{2\omega_{\!\vec{\kappa}}}+\frac{1}{e^{\beta(\omega_{\!\vec{\kappa}}+
\mu)}-1}\frac{e^{\omega_{\!\vec{\kappa}}\tau}}{2\omega_{\!\vec{\kappa}}}\right)
\nonumber \\
     & &
\end{eqnarray}
where $\omega_{\!\vec{\kappa}}=k^{2}+m^{2}$.

The first term of the propagator is the zero-temperature contribution which
shows us the poles. The other two are clearly interpretable as the thermal
contributions. There appear, infact, Bose-Einstein distributions for particles
and anti-particles. Most of all, it is these terms that contain all the
dependencies from the macroscopic parameters which, as one would expect, do not
appear in the poles.
We have infact two poles, in the same position in which we would find them if
in the absence of chemical potential:
\begin{equation}
k_{0}=\pm i \omega_{\!\vec{\kappa}}
\end{equation}
The poles, therefore, do not depend on the macroscopic physics but only on the
form of the microscopic Hamiltonian, the true physical Hamiltonian.

Consider now the propagator for Method I. It can be easily seen that it has two
poles in:
\begin{equation}
k_{0}=i(\pm\omega_{\!\vec{\kappa}}-\mu)
\end{equation}
This is quite a peculiar result as the poles depend on the chemical potential;
we therefore have the presence of a parameter characteristic of ``many-body''
in a microscopic term.
In his work Evans shows how this paradox can be recomposed by studying the
relationship between the fields, in the Heisenberg representation, used in the
two methods. These are not the same due to the different definitions of the
evolution operator. The relationship between two Green functions is such that
the poles are related by a shift of $\mu$ in the energy.

It is important to remember at this time that the two methods are defined by
starting from the same partition function and therefore should provide the same
physical information about the system, even if they can differ in the way they
do this.
Method I, though, requires a much greater attention. The energies of the
various particles are not measured with respect to a common zero but are
traslated, through the choice of $H-\mu Q$, by an amount equal to $q\mu$ with
respect to the standard zero of relativistic field theory and of Method II
itself. This could lead to some confusion, also when, for example, we are
dealing with many charges and symmetry groups, as in the standard model.

\section{Interacting field: standard approach}\renewcommand{\theequation}
{\mbox{\arabic{section}.\arabic{equation}}}\setcounter{equation}{0}

In the background field $\varphi$ approximation, the small disturbances
operator for Method I is:
\begin{equation}
\left(\begin{array}{cc}
{-\partial_{\tau}^{2}-\vec{\nabla}^{2}+m^{2}-\mu^{2}+
\frac{\lambda}{2}\varphi_{1}^{2}+\frac{\lambda}{6}\varphi_{2}^{2}} & 
{-2i\mu\partial_{\tau}+\frac{\lambda}{3}\varphi_{1}\varphi_{2}} \\ 
{2i\mu\partial_{\tau}+\frac{\lambda}{3}\varphi_{1}\varphi_{2}} & 
{-\partial_{\tau}^{2}-\vec{\nabla}^{2}+m^{2}-\mu^{2}+\frac{\lambda}{6}
\varphi_{1}^{2}+\frac{\lambda}{2}\varphi_{2}^{2}} \end{array} \right) 
\end{equation}
It is a matrix valued, non-self-adjoint, elliptic differential operator.

If we compare this with the analogous expression for the free field, we notice
immediately the presence of the terms in $\lambda$ on the off-diagonal. It is
actually the presence of $\mu$ and $\lambda$ together in that position that
complicates things. The eigenvalues of the operators associated with
$\ln\det(A_{I})$ are infact: 
\begin{equation}
\omega_{n}^{2}+k^{2}+m^{2}-\mu^{2}+\frac{\lambda}{3}\varphi^{2}\pm\sqrt{
\frac{\lambda^{2}}{36}\varphi^{4}-4\mu^{2}\omega_{n}^{2}}
\end{equation}
If we did not have the interaction the root would disappear and the same would
occur if there was no net charge. We know, infact, that with Method I problems
arise only for the interacting field. In this state, though, the eigenvalues
cannot be manipulated easily. The standard way of proceeding is to write the
trace of the heat kernel and use the Poisson summation formula. Here, though,
the root is a great obstacle and one does not obtain much. Not even an
approximation for high-temperatures made directly on the eigenvalues produces
useful results. In literature (even if they are calculations on curved spaces)
usually at this point one has to limit oneself to the expansion of the heat
kernel and then proceed by approximation \cite{kirtom}.

The one loop effective potential calculated by Benson, Bernstein and Dodelson
through formal manipulation of infinities is anyway:
\begin{eqnarray}
V_{eff}&=&\frac{1}{2}(m^{2}-\mu^{2})\varphi^{2}+\frac{\lambda}{4!}\varphi^{4}+
\int\frac{d^{3}k}{(2\pi)^{3}} \frac{E_{+}(k)+E_{-}(k)}{2}\nonumber \\
 &+&\frac{1}{\beta}\int\frac{d^{3}k}{(2\pi)^{3}} \ln(1-e^{-\beta
E_{+}(k)})+
\frac{1}{\beta}\int\frac{d^{3}k}{(2\pi)^{3}} \ln(1-e^{-\beta E_{-}(k)})
\end{eqnarray}
with:
\begin{equation}
E_{\pm}(k)^{2}=k^{2}+m^{2}+\mu^{2}+\frac{\lambda}{3}\varphi^{2}\pm\sqrt{
\frac{\lambda^{2}}{36}\varphi^{4}+4\mu^{2}(k^{2}+m^{2}+\frac{\lambda}{3}
\varphi^{2})}
\end{equation}

Benson, Bernstein and Dodelson themselves study only the various limits of this
result. The non-interacting limit coincides with that expected and, in the
limit $\mu$ tending to zero with broken symmetry, one finds Goldstone
excitations. The high-temperature limit turns out to be much more complicated. 

\section{Interacting field: a new approach}\renewcommand{\theequation}
{\mbox{\arabic{section}.\arabic{equation}}}\setcounter{equation}{0}

The use of Method II for the interacting field is not as easy as for the free
field. Due to the new pseudoperiodic conditions on the fields infact, it is not
possible to proceed with the usual expansion with respect to a constant
$\stackrel{\,\_}{\phi}$. {\em The background field $\stackrel{\,\_}{\phi}$ must
be dependent on the imaginary time}. It is infact possible to study the
relationship between the fields in Method I and in Method II \cite{evans} and,
keeping in mind the different definitions of the evolution operator in the two
cases, one obtains:
\begin{equation}
\phi_{II}(\tau)=e^{-\mu\tau}\phi_{I}(\tau)
\end{equation}
where I have labeled the proper fields of each method.
In particular, when one studies a system with symmetry breaking which has a
background field $\stackrel{\,\_}{\phi}\neq 0$:
\begin{equation}
\stackrel{\,\_}{\phi}_{II}(\tau)=e^{-\mu\tau}\stackrel{\,\_}{\phi}_{I}(\tau)=
e^{-\mu\tau}\varphi \label{background}
\end{equation}
with $\varphi$ constant as the background field in Method I is constant. From
now on I will omit the index II. 

Let us see what this dependency implies in the context of an expansion of the
action. If I expand the action up to the second order in the fields with
respect to $\stackrel{\,\_}{\phi}$, I obtain:
\begin{eqnarray}
S\!\!\!&=&\!\! S^{0}+S^{1}+S^{2}=\nonumber \\
 \!\!\!&=&\!\! S(\stackrel{\,\_}{\phi})+\!\int\! d^{4}x\left(\left.
\frac{\delta S}{\delta\phi_{i}(x)}\right|_{\phi=\stackrel{\,\_}{\phi}}
\right)\eta_{i}(x)+\frac{1}{2}\!\int\! d^{4}x\,d^{4}y\;\eta_{i}(x)
\left(\left.\frac{\delta^{2} S}{\delta\phi_{i}(x)\delta\phi_{j}(y)}
\right|_{\phi=\stackrel{\,\_}{\phi}}\right) \eta_{j}(y)\nonumber \\
  & &
\end{eqnarray}
With the necessary functional derivatives and remembering that
$-\vec{\nabla}^{2}\stackrel{\,\_}{\phi}_{i}(\tau)=0$ we obtain that:
\begin{equation}
S^{0}=\int_{0}^{\beta}d\tau \int
d^{3}x\left[\frac{1}{2}\stackrel{\,\_}
{\phi}_{i}(\tau)(-\partial_{\tau}^{2})\stackrel{\,\_}{\phi}_{i}(\tau)+
\frac{1}{2}m^{2}\stackrel{\,\_}{\phi}^{2}(\tau)+\frac{\lambda}{4!}
\stackrel{\,\_}{\phi}^{4}(\tau)\right]
\end{equation}
\begin{equation}
S^{1}=\int_{0}^{\beta}d\tau \int d^{3}x\left(-\partial_{\tau}^{2}
\stackrel{\,\_}{\phi}_{i}(\tau)+m^{2}\stackrel{\,\_}{\phi}_{i}(\tau)+
\frac{\lambda}{6}(\stackrel{\,\_}{\phi}_{1}^{2}(\tau)+\stackrel{\,\_}
{\phi}_{2}^{2}(\tau))\stackrel{\,\_}{\phi}_{i}(\tau)\right)\eta_{i}(x)
\end{equation}
and:
\begin{equation}
S^{2}=\frac{1}{2}\int_{0}^{\beta}d\tau \int d^{3}x\;\eta_{i}(x)\,
A_{i,j}\, 
\eta_{j}(x)
\end{equation}
where $A_{i,j}$ is the matrix that represents the small disturbances operator
$A$:
\begin{equation}
\left(\begin{array}{cc} {-\partial_{\tau}^{2}-\vec{\nabla}^{2}+m^{2}+
\frac{\lambda}{2}\stackrel{\,\_}{\phi}_{1}^{2}+\frac{\lambda}{6}
\stackrel{\,\_}{\phi}_{2}^{2}} & {\frac{\lambda}{3}\stackrel{\,\_}
{\phi}_{1}\stackrel{\,\_}{\phi}_{2}} \\
{\frac{\lambda}{3}\stackrel{\,\_}
{\phi}_{1}\stackrel{\,\_}{\phi}_{2}} &
{-\partial_{\tau}^{2}-\vec{\nabla}^{2}+
m^{2}+\frac{\lambda}{6}\stackrel{\,\_}{\phi}_{1}^{2}+\frac{\lambda}{2}
\stackrel{\,\_}{\phi}_{2}^{2}} \end{array} \right) 
\end{equation}
Let us observe first the operator $A$. It is a matrix valued elliptic
differential operator. It is not self-adjoint, not because of its form, but due
to the boundary conditions, whereas the analogous $A_{I}$ for Method I had
``good'' boundary conditions but the presence of the chemical potential did not
render it self-adjoint anyway.
This operator has a great merit compared to that of Method I: the differential
terms which caused so many problems do not appear on the off-diagonal any
more. It would seem therefore that this approach greatly simplifies the
calculations and the eventual results.
Unfortunately there is a price to pay for this: the term
$\stackrel{\,\_}{\phi}(\tau)$ is also in the small disturbances operator.

With regards to the term $S^{0}$, we would expect to obtain from it the
constant classical potential, something which does not happen if the
$\stackrel{\,\_}{\phi}$ depends on time $\tau$. 

Also the term $S^{1}$ causes us some problems.
I here show fully the equations of motion:
\begin{eqnarray}
\left(-\partial_{\tau}^{2}+m^{2}+\frac{\lambda}{6}(\stackrel{\,\_}
{\phi}_{1}^{2}(\tau)+\stackrel{\,\_}{\phi}_{2}^{2}(\tau))\right)
\stackrel{\,\_}{\phi}_{1}(\tau)=0\nonumber \\
\left(-\partial_{\tau}^{2}+m^{2}+\frac{\lambda}{6}(\stackrel{\,\_}
{\phi}_{1}^{2}(\tau)+\stackrel{\,\_}{\phi}_{2}^{2}(\tau))\right)
\stackrel{\,\_}{\phi}_{2}(\tau)=0
\end{eqnarray}
These are the conditions necessary for the term $S^{1}$ to be cancelled and so
that we can proceed as usual.
It can be easily seen, though, that our
$\stackrel{\,\_}{\phi}(\tau)=e^{-\mu\tau}\varphi$, if it is not constant
(i.e. $\mu = 0$) or null, cannot be a solution to these equations. We would
infact have:
\begin{equation}
\left(-\mu^{2}+m^{2}+\frac{\lambda}{6}(\stackrel{\,\_}{\phi}_{1}^{2}(\tau)+
\stackrel{\,\_}{\phi}_{2}^{2}(\tau))\right)e^{-\mu\tau}\varphi_{i}=0
\end{equation}
and therefore, unless $\varphi$ is zero:
\begin{equation}
(\stackrel{\,\_}{\phi}_{1}^{2}(\tau)+\stackrel{\,\_}{\phi}_{2}^{2}(\tau))=
\frac{6}{\lambda} (\mu^{2}-m^{2})=\mbox{constant}
\end{equation}
The only {\em exact} solution to the equations of motion (for $\mu\neq 0$),
therefore, is obtained for a null background field, i.e. when the simmetry is
unbroken.

I was able to overcome the obstacles described above by resorting to a
high-temperature expansion, which is anyway what we are most interested in as
it is in that range that we want to study phase transition phenomena.
As I showed before:
\begin{equation}
\stackrel{\,\_}{\phi}=e^{-\mu\tau}\varphi=(1-\mu\tau+\frac{1}{2}\mu^{2}
\tau^{2}+\ldots)\varphi
\end{equation}
For high temperatures $\beta$ is small and therefore $\tau$ is also.
We could therefore limit ourselves to substituting $\stackrel{\,\_}{\phi}$ with
its zero order approximation in $\tau$, $\varphi$.
This would seem, at first sight, a good solution to our problems with the
equations of motion. Doing this, however, we eliminate very important physical
information. The tree term of the effective potential would not be dependant on
the chemical potential anymore. Whereas with Method I the chemical potential is
explicitly present in $S^{0}$ as $\frac{1}{2}\mu^{2}\stackrel{\,\_}{\phi}^{2}$,
here it is introduced due to the dynamic term
$-\partial_{\tau}^{2}\stackrel{\,\_}{\phi}$ and therefore would not appear if
we fix $\stackrel{\,\_}{\phi}=\varphi$.

Not even stopping at the second order in the expansion of the exponential is of
any use.
If we obtain the desired term in the dynamic part, there remains, however, the
time dependence in the terms of the type $m^{2}\stackrel{\,\_}{\phi}^{2}$. 

Let us take instead $\stackrel{\,\_}{\phi}(\tau)$ and substitute it in the
equations of motion.We obtain:
\begin{equation}
(-\mu^{2}\varphi_{i}+O(\tau))+m^{2}(1-\mu\tau+\frac{1}{2}\mu^{2}\tau^{2}+
\ldots)\varphi_{i}+\frac{\lambda}{6}\stackrel{\,\_}{\phi}^{2}(\tau)
\stackrel{\,\_}{\phi}_{i}(\tau)=0
\end{equation}
I limit myself to taking the zero order in $\tau$ of the equations of motion:
\begin{equation}
-\mu^{2}\varphi_{i}+m^{2}\varphi_{i}+\frac{\lambda}{6}\varphi^{2}\varphi_{i}=0
\end{equation}
Now a non-zero $\varphi$ can be a solution.
I have not, therefore, done an expansion for high temperatures of the
background field, but of the equations of motion themselves, avoiding in this
way the loss of relevant physical terms and {\em showing how in the
high-temperature approximation the term $S^{1}$ is negligible.}
The same can be done with the zero order term of the action:
\begin{eqnarray}
S^{0}&=&\int_{0}^{\beta}d\tau \int
d^{3}x\left[-\frac{1}{2}\mu^{2}\varphi^{2}+
\frac{1}{2}m^{2}\varphi^{2}+\frac{\lambda}{4!}\varphi^{4}\right]\nonumber \\
     &=&\beta
V[-\frac{1}{2}\mu^{2}\varphi^{2}+\frac{1}{2}m^{2}\varphi^{2}+
\frac{\lambda}{4!}\varphi^{4}] \label{s0intvarphi}
\end{eqnarray}
and with the small disturbances operator:
\begin{equation}
\left(\begin{array}{cc} {-\partial_{\tau}^{2}-\vec{\nabla}^{2}+m^{2}+
\frac{\lambda}{2}\varphi_{1}^{2}+\frac{\lambda}{6}\varphi_{2}^{2}} & 
{\frac{\lambda}{3}\varphi_{1}\varphi_{2}} \\
{\frac{\lambda}{3}\varphi_{1}
\varphi_{2}} & {-\partial_{\tau}^{2}-\vec{\nabla}^{2}+m^{2}+
\frac{\lambda}{6}\varphi_{1}^{2}+\frac{\lambda}{2}\varphi_{2}^{2}} 
\end{array} \right) 
\end{equation}

I will proceed therefore in the calculation in the one loop approximation of
the effective potential using this operator, which, unlike Method I, allows me
to exploit the zeta-function techniques. For the form of the operator we cannot
here exclude a priori the presence of multiplicative anomaly.

We must remember that, in the case of broken symmetry, it is only a
high-temperature approximation. It is, instead, an exact result when the
symmetry is not broken. 

As expected, the eigenvalues of the operators associated with $A$, calculated
in an analogous manner to \ref{loganom}, have a particularly simple form:
\begin{equation}
(\omega_{n}+i\mu)^{2}+k^{2}+m^{2}+\frac{\lambda}{3}\varphi^{2}\pm
\frac{\lambda}{6}\varphi^{2} \label{autovint}
\end{equation}

I will avoid here, for brevity, the explicit calculations. These can be found
in the appendix.
The one loop effective potential, calculated using the approach for the
inclusion of the chemical potential that I developed, results finally as:
\begin{eqnarray}
\!\!\!&\!\!\!&\!\!\!V_{eff}=\frac{S^{0}}{\beta
V}-\frac{\zeta'(0|A)}{2\beta V}+
\frac{\zeta(0|A)}{2\beta V}\ln(\ell^{2})=\nonumber \\
 \!\!\!&\!\!\!&\!\!\!=\frac{1}{2}(m^{2}-\mu^{2})\varphi^{2}+
\frac{\lambda}{4!}\varphi^{4}-\nonumber \\
\!\!\!&\!\!\!&\!\!\!- \frac{1}{32\pi^{2}}\frac{M_{1}^{4}}{2}
\left[-\ln(M_{1}^{2}\ell^{2})+\frac{3}{2}\right]-\frac{1}{32\pi^{2}}
\frac{M_{2}^{4}}{2}\left[-\ln(M_{2}^{2}\ell^{2})+\frac{3}{2}\right]+\nonumber
\\
\!\!\!&\!\!\!&\!\!\!+\!
\frac{1}{2\beta}\!\int\!\frac{d^{3}k}{(2\pi)^{3}} 
\ln(1\!-\!e^{-\beta(\sqrt{k^{2}+M_{1}^{2}}-\mu)})\!+\!\frac{1}{2\beta}\!
\int\!\frac{d^{3}k}{(2\pi)^{3}}
\ln(1\!-\!e^{-\beta(\sqrt{k^{2}+M_{1}^{2}}+
\mu)})\!+\nonumber \\
\!\!\!&\!\!\!&\!\!\!+\!
\frac{1}{2\beta}\!\int\!\frac{d^{3}k}{(2\pi)^{3}}  
\ln(1\!-\!e^{-\beta(\sqrt{k^{2}+M_{2}^{2}}-\mu)})\!+\!
\frac{1}{2\beta}\!
\int\!\frac{d^{3}k}{(2\pi)^{3}}
\ln(1\!-\!e^{-\beta(\sqrt{k^{2}+M_{2}^{2}}+
\mu)})\nonumber \\
           & & \label{mio}
\end{eqnarray}
where $M_{1}^{2}=m^{2}+\frac{\lambda}{2}\varphi^{2}$,
$M_{2}^{2}=m^{2}+\frac{\lambda}{6}\varphi^{2}$.

It was obtained using zeta-function techniques and, as such, is regularised.
It is to be noted that the excitation energies are all particularly clear.

This result naturally includes within itself also the case of non-interacting
field. In the limit of $\lambda$ tending to zero we obtain infact, for the
effective potential, the exact expression:
\begin{eqnarray}
V_{eff}&=&-\frac{1}{16\pi^{2}}\frac{m^{4}}{2}\left[-\ln(m^{2}\ell^{2})+
\frac{3}{2}\right]+\nonumber \\
       &+& \frac{1}{\beta}\int\frac{d^{3}k}{(2\pi)^{3}}  
\ln(1-e^{-\beta(\sqrt{k^{2}+m^{2}}-\mu)})+\frac{1}{\beta} 
\int\frac{d^{3}k}{(2\pi)^{3}}
\ln(1-e^{-\beta(\sqrt{k^{2}+m^{2}}+\mu)})
\nonumber \\
\end{eqnarray}
This expression coincides with that calculable with my approach starting
directly from a free field, in other words using the \ref{tracciadue} for the
\ref{well-known}. As such it coincides, except for the possible multiplicative
anomaly, to the expression calculated in the classical manner and widely
studied in literature. 

We know that this is also an exact calculation for the effective potential when
the symmetry is unbroken. If $\stackrel{\,\_}{\phi}=\varphi = 0$, then
$M_{i}^{2}=m^{2}$ and we easily obtain a coincident expression to that of
Benson, Bernstein and Dodelson's analogous limit.

If we are in conditions of broken symmetry, this expression cannot be seen as
anything other than a high-temperature approximation.

Let us consider the case, with $m^{2}<0$, of symmetry spontaneously broken
even at zero temperature. Now, the value of $\varphi$ that minimises the
tree-level potential is necessarily such that: 
\begin{equation}
\varphi^{2}=\frac{6}{\lambda}(\mu^{2}-m^{2})
\end{equation}
Substituing it in the thermal excitation energies of \ref{mio}, we get:
\begin{eqnarray}
E_{1}(k)=\sqrt{k^{2}+m^{2}+\frac{\lambda}{2}\varphi^{2}}\pm\mu
=\sqrt{k^{2}-
2m^{2}+3\mu^{2}}\pm\mu\nonumber \\
E_{2}(k)=\sqrt{k^{2}+m^{2}+\frac{\lambda}{6}\varphi^{2}}\pm\mu =
\sqrt{k^{2}+
\mu^{2}}\pm\mu 
\end{eqnarray}
Now taking the limit for $\mu$ tending to zero:
\begin{equation}
E_{1}(k)=\sqrt{k^{2}+2|m^{2}|}\;\;\;\;\;\;\;\;\;\;E_{2}(k)=\sqrt{k^{2}}
\end{equation}
As we can see from \ref{background}, for $\mu=0$ the background field is
constant and hence we return to an exact result. We can therefore interpret
these expressions as the energies corresponding to the expected Goldstone boson
and a particle of mass $\sqrt{2|m^{2}|}$.

I would like to point out that the expression obtained by Benson, Bernstein and
Dodelson, though it may have been exact, in its complexity does not allow us to
extract much more information about the system than those obtained with my
result. The authors themselves, must limit themselves to a high-temperature
expansion when studying the phase transition, the influence of the chemical
potential and when giving an evaluation of the critical temperature. Moreover,
their method of calculating this approximation is rather complicated and
requires, for example, the analysis of two different regimes of charge
density. With the expression I have just derived, instead, obtaining the
high-temperature expansion is nearly immediate.

Let us observe, infact, that the thermal terms of \ref{mio} have the same form
as those of the free case:
\begin{equation}
 \frac{1}{\beta}\int\frac{d^{3}k}{(2\pi)^{3}}
\ln(1-e^{-\beta(\sqrt{k^{2}+
m^{2}}-\mu)})+\frac{1}{\beta} \int\frac{d^{3}k}{(2\pi)^{3}}  
\ln(1-e^{-\beta(\sqrt{k^{2}+m^{2}}+\mu)})
\end{equation}
The high-temperature expansion of this expression was first calculated by Haber
and Weldon \cite{hw2}\cite{hw3}. It is well-known and now easily obtainable
using a Mellin-Barnes representation of the zeta function \cite{rew}.
The result is, in the leading terms in $T$:
\begin{equation}
V_{eff}=-\frac{\pi^{2}T^{4}}{45}+\frac{(m^{2}-2\mu^{2})T^{2}}{12}
\end{equation}
If I use the evident analogy between my result and the free case I obtain: 
\begin{eqnarray}
V_{eff}&=&V_{tree}+\frac{1}{2}\left[-\frac{\pi^{2}T^{4}}{45}+
\frac{(M_{1}^{2}-2\mu^{2})T^{2}}{12}-\frac{\pi^{2}T^{4}}{45}+
\frac{(M_{2}^{2}-2\mu^{2})T^{2}}{12}\right]=\nonumber \\
       &=&V_{tree} -\frac{\pi^{2}T^{4}}{90}+\frac{(m^{2}+
\frac{\lambda}{2}\varphi^{2}-2\mu^{2})T^{2}}{24}-\frac{\pi^{2}T^{4}}{90}+
\frac{(m^{2}+\frac{\lambda}{6}\varphi^{2}-2\mu^{2})T^{2}}{24}=\nonumber \\
       &=&V_{tree} -\frac{\pi^{2}T^{4}}{45}+\frac{(2m^{2}+
\frac{\lambda}{2}\varphi^{2}+\frac{\lambda}{6}\varphi^{2}-4\mu^{2})T^{2}}{24}=
\nonumber \\
       &=&
\frac{1}{2}(m^{2}-\mu^{2})\varphi^{2}+\frac{\lambda}{4!}\varphi^{4}
               -\frac{\pi^{2}T^{4}}{45}+\frac{(m^{2}+\frac{\lambda}{3}
\varphi^{2}-2\mu^{2})T^{2}}{12}=\nonumber \\
       &=&
\frac{1}{2}(m^{2}-\mu^{2})\varphi^{2}+\frac{\lambda}{4!}\varphi^{4}
               -\frac{\pi^{2}T^{4}}{45}+\frac{m^{2}T^{2}}{12}+
\frac{\lambda\varphi^{2}T^{2}}{36}-\frac{\mu^{2}T^{2}}{6}
\end{eqnarray}

This result coincides, disregarding $O(T)$ orders in the thermal part, with
that obtained by Benson, Bernstein and Dodelson. The terms in $T^{4},\,T^{2}$
are dominant and one usually limits oneself to these for studying the phase
transition. The difference between Benson, Bernstein and Dodelson's result and
mine appears only from the term in $T$ of the thermal contribution onwards, but
this is reasonable as they are both approximate calculations.

Benson, Bernstein and Dodelson's estimations for the transition temperature and
the interpretation of the spontaneous symmetry breaking as Bose-Einstein
condensation are therefore confirmed by my approach here.

\section{Conclusions}\renewcommand{\theequation}
{\mbox{\arabic{section}.\arabic{equation}}}\setcounter{equation}{0}

I have here analysed the two different methods for the inclusion of the
chemical potential. I showed the existence of some weaknesses of the standard
approach, demonstrating how the development of a new alternative approach to
the problem is not only interesting but necessary for verifying the actual
possibility itself of including the chemical potential in an interacting
theory.

Such an approach was developed using Method II and the zeta-function and
heat-kernel regularisation techniques .

Applied to the case of a non-interacting field I showed that it was equivalent
to the well-known and widely used method if the multiplicative anomaly, which
could complicate the result of Method I, was disregarded. It, instead, does not
appear in my approach which seems here mathematically more rigorous.

I then obtained the propagator and compared it with that calculated with Method
I, showing how the new approach distinguishes more clearly between microscopic
dynamics and boundary conditions.

I then applied the new approach to the more complicated interacting case. A
comparison between the small disturbances operator in the two approaches showed
that the new method was more suitable for a regularised calculation. I obtained
then a new expression for the one loop effective potential for the interacting
field. It is the first expression for this effective potential obtained using
Method II, a method that, as I highlighted and according to other authors, does
not soffer from some of the ambiguities of Method I. The new expression is
regularised but, in the case of broken symmetry, is a high-temperature
approximation. The corresponding excitation energies are also less cryptic than
those obtained by Benson, Bernstein and Dodelson with Method I.

I then studied some relevant limits. In the limit of non-interacting field I
again obtained the expected expression, and likewise in the case of unbroken
symmetry. In the case of symmetry breaking with negative mass squared, the
expected Goldstone excitations appear. Above all, the high-temperature
expansion of my expression turns out to have been obtained in a much simpler
and clearer way than that of Benson Bernstein and Dodelson. My expression
coincides with it, however, in the leading terms in $T$, confirming, therefore,
the results obtained by these authors for the symmetry breaking at high
temperatures. 

This new approach could be further developed. For example, an analysis of the
Seeley-De Witt coefficients could resolve certain unclear aspects of the
regularisation in the presence of chemical potential, it could be applied to
curved spaces and the simpler excitation energies could be of great help in
problems of dynamics of out of equilibrium systems, where knowing the
``masses'' of the particles in play is crucial \cite{rivers}\cite{bolesi}. 

\par\section*{Acknowledgments}
I would like to acknowledge the advice and
contributions of S. Zerbini through all the phases of this investigation. I
would also like to thank R. Rivers and T. Evans for useful correspondence and
comments. 

\renewcommand{\thesection}{\Alph{section}}\setcounter{section}{0}
\section*{Appendix: Explicit calculations}\renewcommand{\theequation}
{\mbox{A.\arabic{equation}}}\setcounter{equation}{0}

To make the discussion more compact I here present the necessary calculations
for obtaining the interacting one loop effective potential using my
approach. Although it could appear complicated, these are standard calculations
of the zeta function technique.
 
I define:
\begin{equation}
M_{1}^{2}=m^{2}+\frac{\lambda}{2}\varphi^{2}\;\;\;\;\;\;\;\;\;M_{2}^{2}=m^{2}+
\frac{\lambda}{6}\varphi^{2}
\end{equation}
Substituting in \ref{autovint}, the trace of the heat kernel is therefore:
\begin{equation}
Tr\,e^{-tA}=V\int\frac{d^{3}k}{(2\pi)^{3}}\sum_{n} 
 [e^{-t[(\omega_{n}+i\mu)^{2}+k^{2}+M_{1}^{2}]}+e^{-t[(\omega_{n}+i\mu)^{2}+
k^{2}+M_{2}^{2}]}]
\end{equation}
Using Poisson's summation formula \cite{rew}:
\begin{equation}
\sum_{k\in {Z\!\!\! Z}^{N}}f(k+q) =\sum_{k\in {Z\!\!\! Z}^{N}}
\int_{{R\!\!\!\! I\,}^{N}} f(x)e^{2\pi i k(x-q)}d^{N}x \label{poisson}
\end{equation}
I can write:
\begin{eqnarray} 
\sum_{n} e^{-t(\omega_{n}+i\mu)^{2}} &=&\sum_{n}
e^{-t\left(\frac{2\pi}{\beta}
\right)^{2}(n+\frac{\beta}{2\pi}i\mu)^{2}} =\nonumber\\
                            &=&\sum_{n}\int_{-\infty}^{+\infty}
e^{-t\left(\frac{2\pi}{\beta}\right)^{2}x^{2}}e^{2\pi i n 
(x-\frac{\beta}{2\pi}i\mu)}dx = \nonumber\\
                            &=&\frac{\beta}{2\sqrt{\pi}}\sum_{n} 
e^{\beta\mu n} t^{-\frac{1}{2}}e^{-\frac{\beta^{2}n^{2}}{4t}} 
\end{eqnarray}
The zeta function will then be:
\begin{eqnarray}
\zeta(s|A)&=&\frac{1}{\Gamma (s)}\int_{0}^{\infty} t^{s-1} Tr\,e^{-tA}
dt=
\nonumber \\
          &=&\frac{\beta
V}{2\sqrt{\pi}}\frac{1}{\Gamma(s)}\int_{0}^{\infty}
t^{s-1-\frac{1}{2}}\int\frac{d^{3}k}{(2\pi)^{3}} 
e^{-t\omega_{\!\vec{\kappa}}^{2}}\sum_{n} e^{\beta\mu n} 
e^{-\frac{\beta^{2}n^{2}}{4t}}dt+\;[M_{1}\rightarrow M_{2}] =\nonumber\\
          &=&\frac{\beta V}{2\sqrt{\pi}}\frac{1}{\Gamma(s)}
\int\frac{d^{3}k}{(2\pi)^{3}} \int_{0}^{\infty}t^{s-\frac{3}{2}} 
e^{-t\omega_{\!\vec{\kappa}}^{2}}dt+\;[M_{1}\rightarrow M_{2}] +\nonumber \\
          &+&\frac{\beta V}{2\sqrt{\pi}}\frac{1}{\Gamma(s)}\sum_{n\geq
1} 
e^{\beta\mu n}\int\frac{d^{3}k}{(2\pi)^{3}} \int_{0}^{\infty}t^{s-
\frac{3}{2}}e^{-t\omega_{\!\vec{\kappa}}^{2}}e^{-\frac{(\beta n 
\omega_{\!\vec{\kappa}})^{2}}{4t\omega_{\!\vec{\kappa}}^{2}}}dt+\;[M_{1}
\rightarrow M_{2}]+\nonumber \\
          &+&\frac{\beta V}{2\sqrt{\pi}}\frac{1}{\Gamma(s)}\sum_{n\geq
1} 
e^{-\beta\mu n}\int\frac{d^{3}k}{(2\pi)^{3}} \int_{0}^{\infty}
t^{s-\frac{3}{2}}e^{-t\omega_{\!\vec{\kappa}}^{2}}e^{-\frac{(\beta n 
\omega_{\!\vec{\kappa}})^{2}}{4t\omega_{\!\vec{\kappa}}^{2}}}dt+\;[M_{1}
\rightarrow M_{2}]\nonumber \\
          & &
\end{eqnarray}
I have distinguished the vacuum term ($n=0$) from the thermal contributions and
have defined $\omega_{\!\vec{\kappa}}^{2}=k^{2}+M_{1}^{2}$. As the two
terms relative to the two eigenvalues differ only in ``mass'', I will only
explicitly indicate one of them.

Let us consider first the vacuum term. Remembering the definition of the
function $\Gamma(s)$:
\begin{eqnarray}
\zeta_{0}(s|A) &=& \frac{\beta V}{2\sqrt{\pi}}\frac{1}{\Gamma(s)} 
\int\frac{d^{3}k}{(2\pi)^{3}}  \int_{0}^{\infty} t^{s-\frac{3}{2}}
e^{-t\omega_{\!\vec{\kappa}}^{2}}\,dt+\;[M_{1}\rightarrow M_{2}] = \nonumber \\
 &=&  \frac{\beta V}{2\sqrt{\pi}}\frac{1}{\Gamma(s)} 
\int\frac{d^{3}k}{(2\pi)^{3}}
(\omega_{\!\vec{\kappa}}^{2})^{\frac{1}{2}-s}
\int_{0}^{\infty} (\omega_{\!\vec{\kappa}}^{2}t)^{s-\frac{3}{2}}
e^{-t\omega_{\!\vec{\kappa}}^{2}}\,d(\omega_{\!\vec{\kappa}}^{2}t)+\;[M_{1}
\rightarrow M_{2}] = \nonumber \\
 &=&  \frac{\beta V}{2\sqrt{\pi}}\frac{1}{\Gamma(s)} 
\int\frac{d^{3}k}{(2\pi)^{3}}  (k^{2}+M_{1}^{2})^{\frac{1}{2}-s}
\Gamma(s-\frac{1}{2})+\;[M_{1}\rightarrow M_{2}]=\nonumber\\
 &=&\frac{\beta
V}{16\pi^{2}}\frac{\Gamma(s-2)}{\Gamma(s)}\,(M_{1}^{2})^{2-s}+
\frac{\beta
V}{16\pi^{2}}\frac{\Gamma(s-2)}{\Gamma(s)}\,(M_{2}^{2})^{2-s} 
\label{zetabesaint}
\end{eqnarray}
In the last step I used $\Gamma(3/2)=\sqrt{\pi}/2$ and the identity:
\begin{equation}
\int_{0}^{\infty}dk\,k^{n-1}(k^{2}+c^{2})^{-z} =
\frac{1}{2}(c^{2})^{n/2-z}
\frac{\Gamma(n/2)}{\Gamma(z)}\Gamma(z-n/2)          \label{envuotoegamma}
\end{equation}
This vacuum contribution is analytic in zero. I can therefore easily
differentiate and obtain:
\begin{equation}
\zeta'_{0}(0|A)=\frac{\beta V}{16\pi^{2}}\frac{M_{1}^{4}}{2}\left[-
\ln(M_{1}^{2})+\frac{3}{2}\right]+\frac{\beta V}{16\pi^{2}}
\frac{M_{2}^{4}}{2}\left[-\ln(M_{2}^{2})+\frac{3}{2}\right]
\end{equation}

I take back two of the terms of the thermal contribution:
\begin{eqnarray}
&&\frac{\beta
V}{2\sqrt{\pi}}\frac{1}{\Gamma(s)}\int\frac{d^{3}k}{(2\pi)^{3}}
\sum_{n\geq 1} e^{\pm\beta\mu n}\int_{0}^{\infty}t^{s-\frac{3}{2}}
e^{-t\omega_{\!\vec{\kappa}}^{2}}e^{-\frac{(\beta n 
\omega_{\!\vec{\kappa}})^{2}}{4t\omega_{\!\vec{\kappa}}^{2}}}dt=\nonumber \\
&=&\frac{\beta
V}{2\sqrt{\pi}}\frac{1}{\Gamma(s)}\int\frac{d^{3}k}{(2\pi)^{3}}
\sum_{n\geq 1} e^{\pm\beta\mu n}
(\omega_{\!\vec{\kappa}}^{2})^{\frac{1}{2}-s}
\int_{0}^{\infty} y^{s-\frac{3}{2}} e^{-y} e^{-\frac{\beta^{2}
\omega_{\!\vec{\kappa}}^{2}n^{2}}{4y}}\,dy
\end{eqnarray}
I use the representation \cite{graryz} of McDonald's functions:
\begin{equation}
K_{\nu}(z) =\frac{1}{2}\left(\frac{z}{2}\right)^{\nu}
\int_{0}^{\infty}
\frac{e^{-t-\frac{z^{2}}{4t}}}{t^{\nu+1}}dt
\end{equation}
With $z=\beta n \omega_{\!\vec{\kappa}}$ e $\nu=1/2-s$:
\begin{equation}
\frac{\beta
V}{2\sqrt{\pi}}\frac{1}{\Gamma(s)}\int\frac{d^{3}k}{(2\pi)^{3}}  
\sum_{n\geq 1} e^{\pm\beta\mu n}    2^{\frac{3}{2}-s}(\beta n
)^{s-\frac{1}{2}}
\omega_{\!\vec{\kappa}}^{\frac{1}{2}-s}K_{\frac{1}{2}-s}
(\beta n \omega_{\!\vec{\kappa}})
\end{equation}
It is an analytic function.
I obtain, therefore, for one of the two cases:
\begin{equation}
\frac{\beta V}{2\sqrt{\pi}}\int\frac{d^{3}k}{(2\pi)^{3}}  \sum_{n\geq
1} 
e^{\beta\mu n} 2^{\frac{3}{2}}(\beta n )^{-\frac{1}{2}}
\omega_{\!\vec{\kappa}}^{\frac{1}{2}}K_{\frac{1}{2}}(\beta n 
\omega_{\!\vec{\kappa}})
\end{equation}
Remembering another useful representation of McDonald's functions
\cite{graryz}:
\begin{equation}
K_{\frac{1}{2}}(z) = \sqrt{\frac{\pi}{2z}}e^{-z} 
\end{equation}
we then pass easily to:
\begin{equation}
 V \int\frac{d^{3}k}{(2\pi)^{3}}  \sum_{n\geq 1}\frac{\left(
e^{-\beta(\omega_{\!\vec{\kappa}}-\mu)}\right)^{n}}{n} 
\end{equation}
And from here to:
\begin{equation}
  - V \int\frac{d^{3}k}{(2\pi)^{3}}
\ln(1-e^{-\beta(\sqrt{k^{2}+M_{1}^{2}}-
\mu)})
\end{equation}
The condition for convergence of the sum is here turned into:
\begin{equation}
\left|e^{-\beta(\omega_{\!\vec{\kappa}}-\mu)}\right|<
1\;\;\Rightarrow\;
\omega_{\!\vec{\kappa}}-\mu>0\;\;\Rightarrow\;\mu<M_{1}
\end{equation}
The term with the negative chemical potential, proceeding analogously, results
in:
\begin{equation}
  - V \int\frac{d^{3}k}{(2\pi)^{3}}
\ln(1-e^{-\beta(\sqrt{k^{2}+M_{1}^{2}}+
\mu)}) 
\end{equation}
with the condition: $-\mu<M_{1}$.

We must remember, though, that we have two terms in $M_{2}$ that give analogous
contributions. By definition, $M_{2}<M_{1}$ and therefore the condition for
summation this time requires that the chemical potential is:
\begin{equation}
|\mu|<\sqrt{m^{2}+\frac{\lambda}{6}\varphi^{2}}
\end{equation}
I therefore found that:
\begin{eqnarray} 
\!\!\!&\!\!\!&\!\!\!\zeta'(0|A)=\frac{\beta
V}{16\pi^{2}}\frac{M_{1}^{4}}{2}
\left[-\ln(M_{1}^{2})+\frac{3}{2}\right]+\frac{\beta V}{16\pi^{2}}
\frac{M_{2}^{4}}{2}\left[-\ln(M_{2}^{2})+\frac{3}{2}\right]-\nonumber \\
 \!\!\!&\!\!\!&\!\!\! - V \int\frac{d^{3}k}{(2\pi)^{3}}  
\ln(1-e^{-\beta(\sqrt{k^{2}+m^{2}+\frac{\lambda}{2}\varphi^{2}}-\mu)})-
V 
\int\frac{d^{3}k}{(2\pi)^{3}}  \ln(1-e^{-\beta(\sqrt{k^{2}+m^{2}+
\frac{\lambda}{2}\varphi^{2}}+\mu)})-\nonumber \\
 \!\!\!&\!\!\!&\!\!\!- V \int\frac{d^{3}k}{(2\pi)^{3}}  
\ln(1-e^{-\beta(\sqrt{k^{2}+m^{2}+\frac{\lambda}{6}\varphi^{2}}-\mu)})-
V 
\int\frac{d^{3}k}{(2\pi)^{3}}  \ln(1-e^{-\beta(\sqrt{k^{2}+m^{2}+
\frac{\lambda}{6}\varphi^{2}}+\mu)})\nonumber \\
 \!\!\!&\!\!\!&\!\!\!\label{zetaprimo0intvarphi}
\end{eqnarray}
To determine the effective potential we need also $\zeta(0|A)\ln(\ell^{2})$.
As all the elements of the thermal contribution are of the form
$f(s)/\Gamma(s)$, these do not contribute as $f(s)$ is analytic and therefore
finite in zero, whereas $\Gamma(s)$ has a pole in zero.
Therefore, remembering that $\zeta_{0}(s|A)$ (\ref{zetabesaint}) is analytic in
zero:
\begin{equation}
\zeta(0|A)\ln(\ell^{2})=\zeta_{0}(0|A)\ln(\ell^{2})= \frac{\beta
V}{16\pi^{2}}
\frac{M_{1}^{4}}{2}\ln(\ell^{2})+\frac{\beta
V}{16\pi^{2}}\frac{M_{2}^{4}}{2}
\ln(\ell^{2})\label{zeta0intvarphi}
\end{equation}
Combining these contributions we obtain my expression \ref{mio}.

\end{document}